# Ionic Diffusion Properties of Rare-Earth High-Entropy Oxides from a Machine-Learned Interatomic Potential

Mary Kathleen Caucci[1], Billy E. Yang[2], Saeed S.I. Almishal[2], Jon-Paul Maria[2], Susan B. Sinnott[*1,2,3,4]

[1]*Department of Chemistry, The Pennsylvania State University, University Park, PA 16802*
[2]*Department of Materials Science and Engineering, The Pennsylvania State University, University Park, PA 16802*
[3]*Materials Research Institute, The Pennsylvania State University, University Park, PA, 16802*
[4]*Institute for Computational and Data Science, The Pennsylvania State University, University Park, PA, 16802*

**ABSTRACT**

Rare-earth high-entropy oxides (RE-HEOs) have emerged as a promising class of functional ceramics for solid-state electrochemical applications due to their chemical complexity, structural tunability, and potential for fast oxygen-ion transport. In this work, we investigate oxygen diffusion in ceria-based RE-HEOs of the form $Ce_x(YLaPrSm)_{1-x}O_{2-\delta}$ using classical molecular dynamics simulations driven by the Crystal Hamiltonian Graph Neural Network (CHGNet) machine-learned interatomic potential. To improve predictive accuracy for lanthanide-containing systems, we benchmark three CHGNet variants, including a fine-tuned r²SCAN-trained model, against targeted density functional theory (DFT) data that explicitly include *f*-valence electrons. Simulations across temperature, Ce content, oxygen vacancy concentration, and both fluorite and bixbyite structures reveal that oxygen transport in RE-HEOs is governed by the interplay of two factors: the concentration of mobile vacancies and the local cation environment through which they hop. At fixed composition, ionic conductivity exhibits a non-monotonic dependence on vacancy concentration, with optimal diffusion occurring at moderate vacancy levels and reduced mobility at higher concentrations. Increasing Ce content lowers migration activation energies and enhances diffusivity through low-barrier diffusion networks built from Ce–Ce and Ce–Y edges. Analysis of individual oxygen hopping events provides atomistic insight into how local chemical environments and short-range cation ordering govern transport in high-entropy oxides. Overall, this work demonstrates that machine-learned interatomic potentials can resolve composition–structure–transport relationships in chemically complex oxides, and identifies Ce–Ce and Ce–Y edges as the active pathways through which compositional tuning can enhance oxygen-ion conductivity in rare-earth high-entropy materials.

**Keywords**
Molecular Dynamics; diffusion; high-entropy oxides

## 1. INTRODUCTION

Ceria based rare-earth high-entropy oxides (RE-HEOs), composed of multiple rare-earth (RE) cations in equimolar or near-equimolar ratios, represent a promising class of functional ceramics that are characterized by high configurational complexity, enhanced thermodynamic stability, and potential for fast oxygen-ion transport. High-entropy oxides were first demonstrated as entropy-stabilized single-phase ceramics that derived their stability from the large configurational entropy associated with multiple cation species occupying a common sublattice [1,2]. In fluorite-structured ceria systems, substitution with multiple rare-earth cations introduces oxygen vacancies $V_O$ and local lattice distortions that can significantly modify oxygen-ion transport behavior [3–11]. As a result, RE-HEOs have attracted increasing interest for solid-state electrochemical applications such as solid oxide fuel cells, oxygen separation membranes, and thermochemical energy conversion systems, where high oxygen ionic conductivity is critical for device efficiency and durability [12–14].

Oxygen diffusion in simpler binary rare-earth oxides (REOs), including such $CeO_2$-based solid solutions doped with Gd, Sm, or Pr, has been studied extensively through both experimental and computational approaches [11,13,15–18]. These studies have established vacancy-mediated hopping as the dominant transport mechanism and have highlighted the sensitivity of migration barriers to dopant size, charge state, and local coordination environment. However, the oxygen transport behavior in RE-HEOs remains largely uncharacterized. The complexity of these systems arising from the interplay of cation disorder, oxygen non-stoichiometry, and local structural heterogeneity, poses a challenge for both experimental characterization and first-principles simulation [19–22]. Notably, the degree to which high configurational entropy disrupts or enhances oxygen vacancy ordering, and whether this helps maintain high-symmetry conductive phases (e.g., disordered fluorite), continues to be an active area of research [23–28].

In this work, we focus on ceria-based RE-HEOs of the form $\mathrm{Ce_x(YLaPrSm)_{1-x}O_{2-\delta}}$, where $x$ controls the relative Ce content and $\delta$ denotes the degree of oxygen non-stoichiometry arising from oxygen vacancies, $V_O$. RE-HEOs are known to crystalize predominantly in either disordered fluorite-type structures or oxygen-deficient bixbyite-derived structures, with phase selection governed by average cation valence, preferred coordination environments, oxygen vacancy concentration, and synthesis conditions [4,26–30]. The fluorite structure, exemplified by $CeO_2$, stabilizes tetravalent cations and is well-known for its ability to accommodate a high degree of oxygen vacancy disorder, underpinning its high oxygen ionic conductivity. In contrast, bixbyite structures are typically preferred by trivalent RE cations, exhibit distorted sixfold cation coordination, and possess ordered oxygen vacancy sublattices that can impede long-range oxygen transport. Prior studies have demonstrated that Ce content plays a critical role in determining structural stability in the $Ce_x(YLaPrSm)_{1-x}O_{2-\delta}$ system, with decreasing Ce concentration driving a transition toward bixbyite-dominated or multiphase materials [4,28,30,31]. This compositional space therefore provides a valuable platform for systematically examining oxygen diffusion across both fluorite and bixbyite structural motifs within a single chemically complex oxide family.

To investigate oxygen diffusion in these systems, we employ classical molecular dynamics (MD) simulations using the Crystal Hamiltonian Graph Neural Network (CHGNet) machine-learned interatomic potential [32]. CHGNet is pre-trained on a broad set of density functional theory (DFT) data from the Materials Project (MP) database [33], enabling longer-timescale simulations with reduced computational cost while promising physically-grounded accuracy. This approach allows us to explicitly assess how cation composition, oxygen vacancy concentration, and crystal structure collectively influence oxygen mobility in chemically and structurally complex RE-HEOs.

To bridge atomistic transport mechanisms with macroscopic transport behavior, we track individual oxygen hopping events throughout the simulations. Oxygen diffusion in fluorite-type oxides occurs primarily through a vacancy-mediated hopping mechanism, where an oxygen ion hops into a neighboring vacancy site [34–37]. In RE-HEOs, these hops proceed through cation-cation (RE-RE) edges, forming the chemically diverse six-atom local hopping environment that consists of the initial and final oxygen sites, the vacancy, and the four surrounding RE cations (Figure 1). The associated migration activation energy barrier, $E_a^{ion}$, is strongly influenced by the RE-RE edge located at the transition point, reflecting the highly heterogeneous local energy landscape intrinsic to disordered systems. This complex, chemically diverse local structure plays a key role in determining the probability and energetics of individual hopping events.

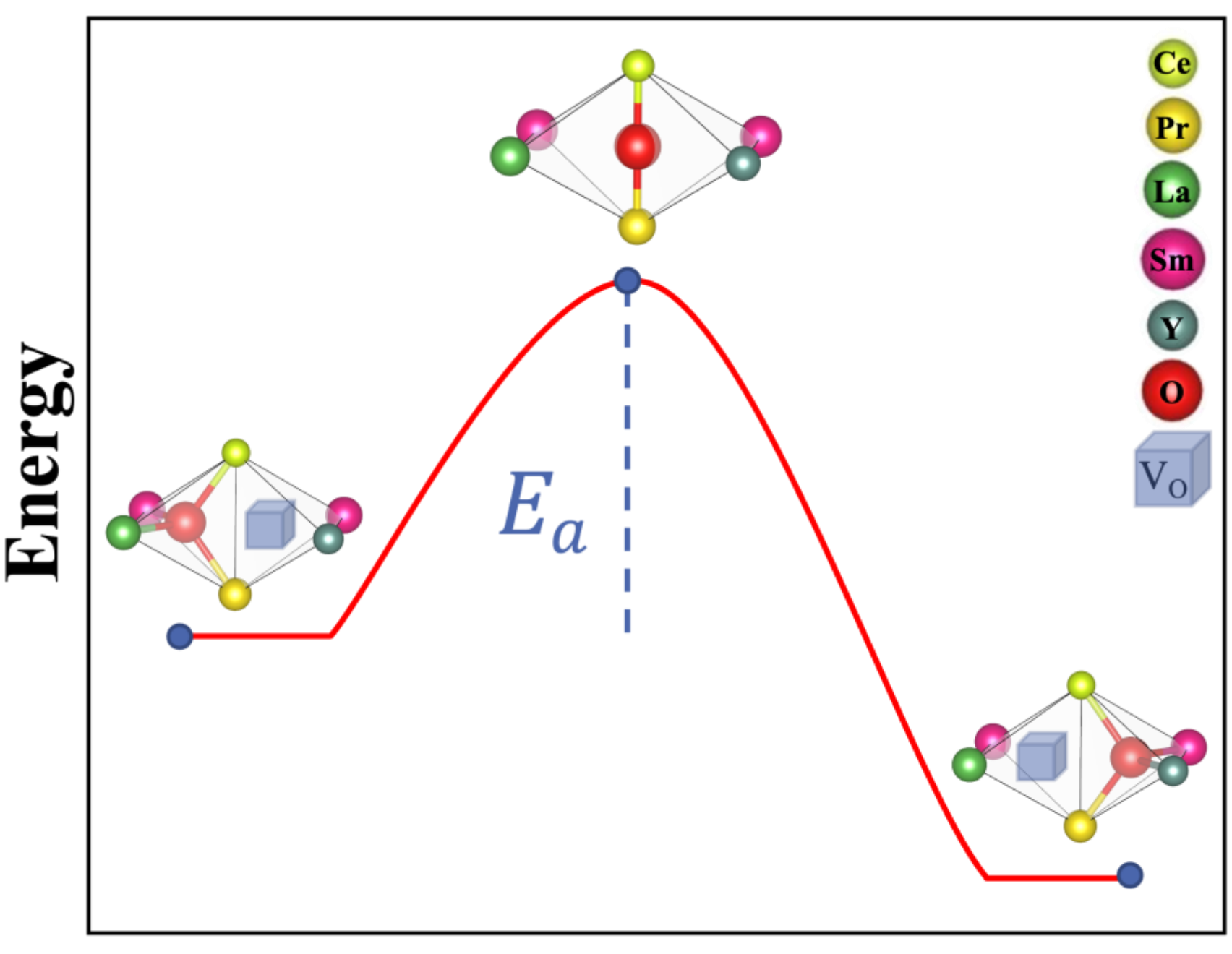

Figure 1. Schematic of vacancy-mediated oxygen migration in the RE-HEO, showing the energy profile along the reaction coordinate (red curve) and the six-atom local environment at three points along the pathway: the initial configuration (left), the transition state at the Ce-Pr migration edge (center), and the final configuration after the hop (right). The activation energy $E_a^{ion}$ is the barrier between the initial and transition states. Atom colors follow the legend (Ce, Pr, La, Sm, Y, O, and the oxygen vacancy $V_O$).

Our aim is to quantify the extent to which high configurational entropy can suppress long-range vacancy ordering, commonly associated with phase transitions to low-symmetry, low-conductivity structures, and instead promote the formation of diffusion networks that enhance long-range oxygen transport [23]. By evaluating temperature-dependent diffusivities and hopping statistics across different compositions, oxygen vacancy concentrations, and crystal structures, we assess the capacity of RE-HEOs to support robust ionic conductivity over a wide range of conditions. We find that, at fixed composition, ionic conductivity depends non-monotonically on vacancy concentration, with optimal transport at moderate vacancy levels and reduced mobility at higher concentrations consistent with vacancy interactions and trapping. Increasing Ce content systematically lowers migration activation energies and enhances diffusivity, driven by the emergence of low-barrier pathways dominated by Ce–Ce and Ce–Y cation–cation edges. Analysis of individual hopping events further reveals how local chemical environments and short-range cation ordering govern transport in these chemically complex oxides. The predicted conductivities are further compared with experimental measurements carried out on the same compositions.

## 2. METHODS

### 2.1. Computational details

Molecular dynamics simulations were performed using the Large-scale Atomic/Molecular Massively Parallel Simulator (LAMMPS) code [38], utilizing the machine-learned CHGNet interatomic potential integrated to LAMMPS [32,39]. The original CHGNet version 0.3.0 was trained on DFT data that employed GGA and GGA+$U$ approximations. More recently, a new CHGNet model was released that was trained on a $r^2$SCAN dataset parsed from the MP database. This CHGNet–$r^2$SCAN model was initialized from the GGA/GGA+$U$-trained CHGNet v0.3.0 weights and subsequently fine-tuned on a dataset computed with the $r^2$SCAN exchange–correlation functional [40]. Although the CHGNet–$r^2$SCAN model was trained on higher-fidelity DFT data compared to CHGNet–v0.3.0, both pre-trained models employed *f*-core pseudopotentials for Pr, Sm, and Nd, in which the partially filled 4*f* electrons are frozen into the pseudopotential core and excluded from the valence manifold. This treatment reduces computational cost and avoids the convergence difficulties associated with partially occupied *f*-states, but at the expense of explicitly capturing *f*-electron contributions to bonding and magnetism [41]. Since both CHGNet–v0.3.0 and CHGNet–$r^2$SCAN exclude *f*-valence electrons for Pr, their predictions may exhibit inaccuracies for systems where explicit treatment of the *f*-electron manifold may be important for modeling diffusion. To address this limitation, we fine-tune the CHGNet–$r^2$SCAN model using our own $r^2$SCAN-based DFT dataset, termed CHGNet–tuned. Our DFT dataset includes the binary rare-earth oxides, rare-earth metals, and RE-HEOs presented in [41,42]. The full DFT dataset used for fine-tuning is openly available as JSON-formatted training files at the Open Science Framework under the DOI https://doi.org/10.17605/OSF.IO/W38CM. The fine-tuning procedure followed the same transfer learning strategy and training configuration described in [40], using the CHGNet–$r^2$SCAN model as the initialization point. All results presented use CHGNet–v0.3.0, unless noted otherwise.

The starting configurations for all simulations of $\mathrm{Ce_x(YLaPrSm)_{1-x}O_{2-\delta}}$ were generated using the special quasi-random structure (SQS) algorithm [43] via the integrated cluster expansion toolkit (ICET) [44]. Four 32-cation SQSs were generated from size 2×2×2 conventional fluorite ($Fm\bar{3}m$) unit cells of $\mathrm{Ce_x(YLaPrSm)_{1-x}O_2}$ with Ce concentrations x = 0.22, 0.31, 0.50, and 0.81 and several oxygen non-stoichiometries δ. For each Ce concentration, an identical cation configuration was constructed to generate the corresponding bixbyite ($Ia\bar{3}$) supercells. In fluorite structures, oxygen vacancies, $V_O$, were introduced randomly, whereas in bixbyite structures the oxygen vacancies were restricted to the Wyckoff *8b* positions, in accordance with the crystallographic $Ia\bar{3}$ site symmetry. Simulations were conducted on equimolar (22% Ce) RE-HEOs with oxygen vacancy concentrations ranging from 15% to 22%, and on Ce-enriched compositions with 31%, 50%, and 81% Ce content containing oxygen vacancies of 14-17%, 10-14%, and 3-9%, respectively. To compare finite-size effects on observed trends, two additional larger 768-atom fluorite SQS were generated with size 4×4×4 conventional fluorite unit cell with 20% and 50% Ce content. These larger supercells for 20% and 50% Ce content both have a total of 256 cations containing 2%-20% and 10%-20% oxygen vacancies, respectively. All RE-HEO results presented use the 32-cation simulations, unless noted otherwise.

Each configuration was simulated at five temperatures (800 K, 975 K, 1000 K, 1200 K, and 1500 K) to capture temperature-dependent diffusion behavior. Simulations followed a three-step protocol: a 150 ps NPT equilibration step to allow the cell volume to relax under finite-temperature and pressure conditions; a 500 ps NVT equilibration at the same temperature using the averaged volume from the NPT step; and a production run of at least 5 ns under the NVT ensemble. A time step of 2 fs (0.002 ps) was used throughout. The Nose–Hoover thermostat and barostat were employed for temperature and pressure control, as implemented in LAMMPS. Trajectory data was saved every 0.02 ps during NPT equilibration, every 0.2 ps during NVT equilibration, and every 2 ps during the production run. For systems exhibiting low oxygen mobility, simulation times were extended up to 20 ns to ensure sufficient sampling and to achieve a mean squared displacement (MSD) of at least 10 Å$^2$ for oxygen atoms. Simulations were terminated if the MSD did not exceed 5 Å$^2$ within 10 ns.

Oxygen diffusion coefficients were extracted from the slope of the oxygen MSD. The MSD of *N* diffusing oxygen atoms was computed using the standard expression:

$$\mathrm{MSD}(t) = \frac{1}{N}\sum_{i=1}^{N}[r_i(t) - r_i(0)]^2 \quad (1)$$

where $r_i(t)$ is the position of atom *i* at time *t*, and $r_i(0)$ is its position at the start of the trajectory. The diffusion coefficient *D* was then obtained from the Einstein relation:

$$D = \frac{MSD(\Delta t)}{2d\Delta t} \tag{2}$$

where $d$ is the dimensionality (here, $d$ = 3). The ionic conductivity σ was estimated from the calculated diffusion coefficient using the Nernst-Einstein relation:

$$\sigma = \frac{Nq^2 D}{V k_{\mathrm{B}} T} \tag{3}$$

where V is the simulation cell volume, $N$ is the number of mobile charge carriers (oxygen vacancies, $N_V$), $q$ is the elementary charge, $k_B$ is the Boltzmann constant, and $T$ is the absolute temperature. The Nernst-Einstein relation can be applied to both mobile ions and vacancies; in this work we interpret transport in terms of a vacancy-mediated hopping mechanism and take $N = N_V$ [6,45,46]. Finally, the temperature dependence of ionic conductivity was analyzed by fitting to the Arrhenius equation:

$$\ln(\sigma T) = \frac{-E_a}{k_{\mathrm{B}} T} + \ln(\sigma_0) \tag{4}$$

where $E_a$ is the activation energy and $\sigma_0$ is the pre-exponential factor corresponding to the high-temperature limit of conductivity. Because the simulated conductivity reflects oxygen-ion migration alone whereas experimentally measured conductivity additionally contains an electronic contribution in these mixed ionic–electronic conductors, we distinguish the two throughout by denoting the activation energy extracted from simulation as $E_{\mathrm{a}}^{\mathrm{ion}}$ and that extracted from experimental total conductivity as $E_{\mathrm{a}}^{\mathrm{tot}}$.

Oxygen ion hops presented in Section 3.3 were identified from the molecular dynamics trajectory by assigning each oxygen atom to its nearest site in an equilibrated reference structure at every frame using a KD-tree; a change in site assignment accompanied by a displacement exceeding 2.25 Å between successive frames was recorded as a hop event. Each hop was attributed to the pair of rare-earth cations nearest the geometric midpoint of the hop vector (computed using the minimum image convention), and hop frequencies were normalized by the static count of each RE–RE pair type present in the reference structure to enable direct comparison across chemically distinct pair environments. A representative analysis script used to perform hop identification and pair attribution is openly available at the Open Science Framework under the DOI https://doi.org/10.17605/OSF.IO/W38CM.

## 3. RESULTS & DISCUSSION

## 3.1. Benchmarking CHGNet

CHGNet is a machine learned interatomic potential (MLIP) trained on the MPtrj dataset, which consists of a large number of static calculations and structural optimization trajectories from the Materials Project database [33]. Although CHGNet captures a wide chemical and structural domain, it was not explicitly trained on multicomponent HEOs. The training data primarily comprises ordered structures with fewer than 100 atoms [47], and does not include disordered fluorite or bixbyite phases. Although our previous work demonstrated CHGNet produces highly accurate predictions for rocksalt structured HEOs [48], benchmarking is necessary to assess its predictive capability for RE-HEO systems.

To evaluate CHGNet's accuracy, we first tested its performance on the same suite of 15 binary REOs previously examined in Ref. [41]. Since our DFT calculation parameters differ from those used in the MPtrj dataset [33,49,50], we cannot directly compare our DFT energies with those from CHGNet; instead, we compare the magnitude of errors from predictions compared to experiment. Equilibrium volumes [Figure 2(a)] and formation enthalpies $\Delta H_f$ [Figure 2(b)] predicted by the three CHGNet models, v0.3.0, CHGNet–r²SCAN, and the fine-tuned CHGNet–r²SCAN (hereafter CHGNet–tuned), are compared against experimental values and results obtained from DFT exchange-correlation functionals from Ref. [41], including PBEsol, and r²SCAN.

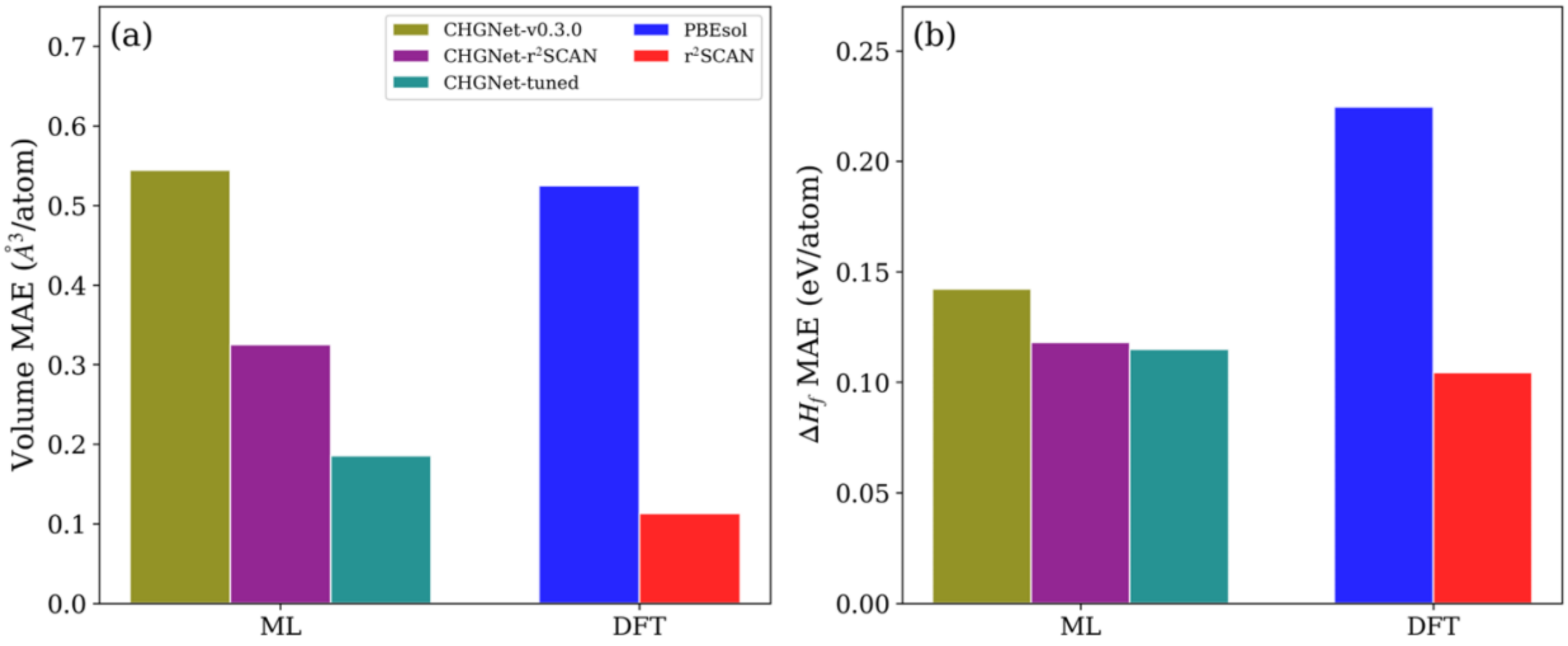

Figure 2. The mean absolute errors, MAEs, of the MLIP and DFT (a) lattice volume predictions and (b) formation enthalpies $\Delta H_f$ compared to experiment for 15 REOs using CHGNet (v0.3.0, r²SCAN, tuned), PBEsol, and r²SCAN.

CHGNet–v0.3.0 demonstrates high fidelity in predicting formation enthalpies, achieving mean absolute errors (MAEs) and mean relative errors (MREs) below 150 meV/atom and 2%, respectively. However, its prediction of equilibrium volume proved less accurate with volume MAE of 0.544 Å$^3$ (3.16% MRE), slightly exceeding those of PBEsol. The newer CHGNet–r$^2$SCAN model, trained on higher-fidelity r$^2$SCAN data, exhibits notable improvements, reducing the volume MAE to 0.325 Å$^3$ (1.81% MRE) and the formation enthalpy MAE to 0.118 eV/atom (–0.65% MRE). Further refinement through fine-tuning the CHGNet–r$^2$SCAN model on our r$^2$SCAN DFT dataset yields the CHGNet–tuned model, which achieves the highest accuracy, with a volume MAE of 0.185 Å$^3$ (–0.04% MRE) and a formation enthalpy MAE of 0.115 eV/atom (–2.25% MRE). These results demonstrate that targeted fine-tuning on system-specific DFT data improves CHGNet's quantitative agreement with both experiment and high-level DFT, particularly in capturing subtle structure–energy relationships among the rare-earth oxides (see also Figure S.1).

MAEs from the pre-trained CHGNet models for both $\Delta H_f$ and volume were particularly larger for compounds involving Pr and Sm, such as $PrO_2$ [Figure S.2]. This limitation stems primarily from CHGNet's lack of training on DFT data that explicitly incorporates *f*-electron physics. In particular, the MPTrj dataset used for training CHGNet–v0.3.0 and CHGNet–r$^2$SCAN employed pseudopotentials in which the 4*f* electrons of Pr, Sm, and Nd were treated as part of the ionic core. As a result, these pre-trained models fail to capture *f*-electron–related magnetic moments, as reflected by their prediction of 0 $\mu_B$ magnetic moments for these species [Figure S.3]. By contrast, the fine-tuned CHGNet–r$^2$SCAN model (CHGNet–tuned), trained on our r$^2$SCAN DFT dataset that includes *f*-valence electrons for Pr, Nd, and Sm, restores nonzero magnetic moments and better reproduces the energetic and structural trends associated with multivalent and magnetically active lanthanides. Consequently, the tuned model mitigates the underestimation of *f*-electron–driven effects present in the pre-trained CHGNet variants, leading to improved accuracy for systems such as RE-HEOs where elements like Pr can exhibit variable oxidation states (e.g., $Pr^{3+}/Pr^{4+}$).

To further assess model performance in multicomponent systems, CHGNet was applied to the same equimolar RE-HEO compositions studied with r$^2$SCAN DFT in Ref. [42]. Figure 3 compares the formation enthalpies of fluorite and bixbyite phases as a function of oxygen non-stoichiometry δ. Consistent with the binary oxide benchmarks, both CHGNet–v0.3.0 and CHGNet–r$^2$SCAN reproduce the overall shape of the $\Delta H_f$ curve but systematically predict more negative formation energies than DFT–r$^2$SCAN. Notably, the CHGNet–r$^2$SCAN results closely follow the same curvature as CHGNet–v0.3.0, though with slightly higher $\Delta H_f$ values that more closely align with DFT predictions, reflecting the influence of the higher-fidelity r$^2$SCAN training dataset.

In terms of phase stability, both pre-trained CHGNet models favor the bixbyite phase beginning at approximately 10% $V_O$, whereas DFT–r$^2$SCAN predicts near-degenerate stability between fluorite and bixbyite up to roughly 15% $V_O$. The fine-tuned CHGNet–r$^2$SCAN (CHGNet–tuned) most closely reproduces the DFT energetics, accurately capturing the relative energy scale and curvature of the $\Delta H_f$ trend. Yet it still slightly over-stabilizes bixbyite at 10% $V_O$, suggesting that while fine-tuning improves agreement with DFT, residual discrepancies may stem from CHGNet's limited description of defect-induced local lattice distortions or its reduced sensitivity to subtle electronic contributions that stabilize the fluorite phase at intermediate vacancy concentrations.

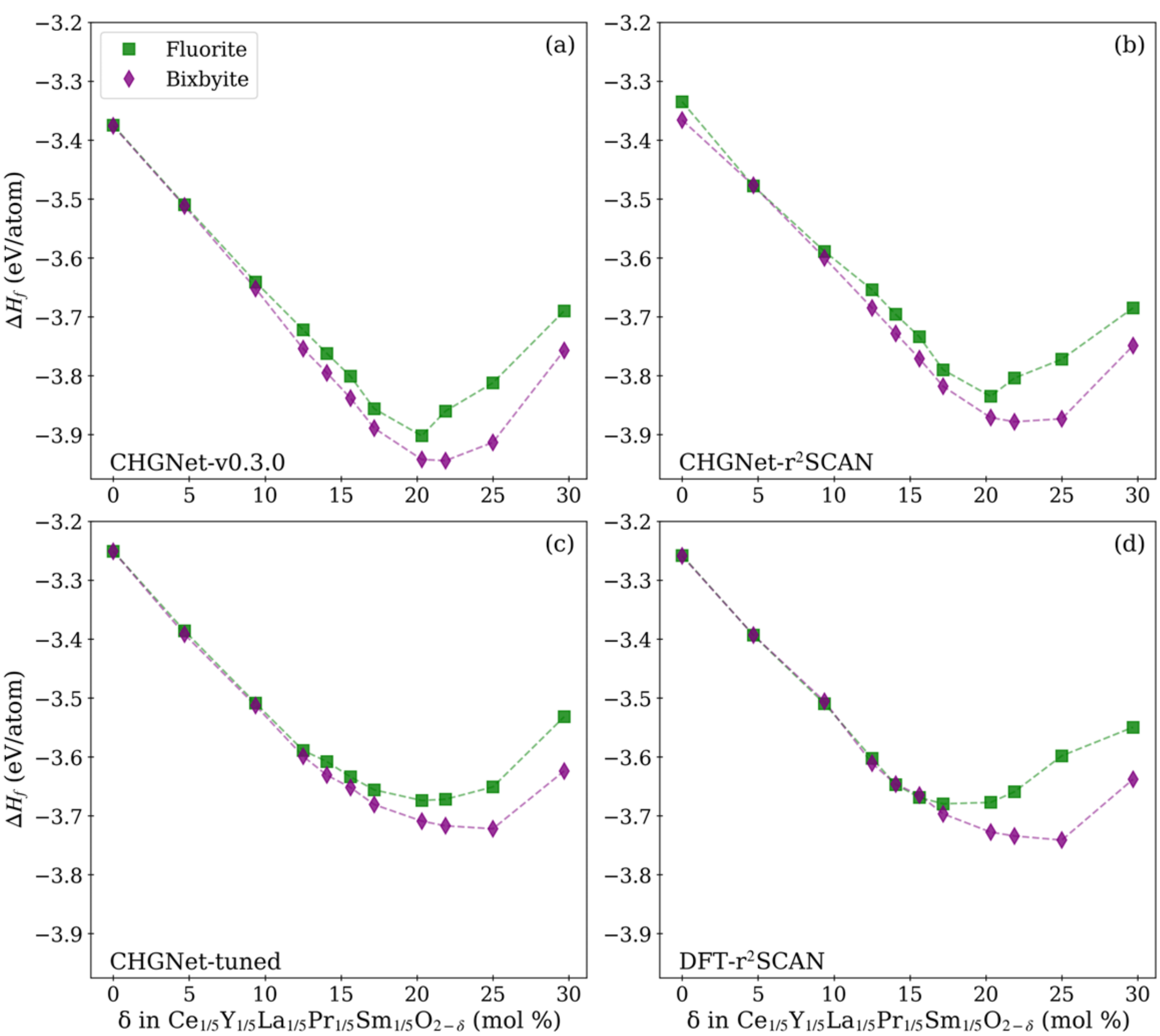

Figure 3. Minimum formation enthalpy $\Delta H_f$ in eV/atom of $Ce_{1/5}Y_{1/5}La_{1/5}Pr_{1/5}Sm_{1/5}O_{2-\delta}$ as a function of oxygen vacancy concentration predicted from either (a) CHGNet-v0.3.0, (b) CHGNet-r$^2$SCAN, (c) CHGNet-tuned, and (d) DFT-r$^2$SCAN reproduced from Ref. [41]. Green squares represent fluorite structures, and purple diamonds represent bixbyite structures. Dotted lines connecting points are to guide the eye.

Overall, CHGNet demonstrates a strong capability for capturing structural and energetic trends in RE-HEOs, with promise to extend our predictive ability to length-scales where DFT is computationally prohibitive. Nonetheless, limitations stemming from its training domain, particularly the neglect of *f*-electron behavior and multivalent cation chemistry, must be carefully considered when interpreting properties that depend on oxidation state flexibility or subtle electronic effects. The CHGNet–r$^2$SCAN model exhibited improved relative errors and closer agreement with r$^2$SCAN DFT data compared to CHGNet–v0.3.0, revealing the benefit of higher-fidelity training data. Further fine-tuning of CHGNet–r$^2$SCAN on our r$^2$SCAN DFT dataset, which explicitly includes *f*-electrons in the valence for Pr, Nd, and Sm, yields the most accurate results overall, achieving the lowest MAE against experimental binary oxide data and most faithfully reproducing the DFT formation-enthalpy trends across composition for the equimolar RE-HEO system. We emphasize that the benchmarks above establish CHGNet's quantitative reliability for the structural and energetic descriptors most relevant to oxygen transport — lattice volume, formation enthalpy, and relative phase stability across the fluorite–bixbyite composition space — even where absolute formation energies and *f*-electron–derived properties carry residual error. The implications of these residual errors for the diffusion predictions reported in subsequent sections are discussed where relevant, particularly in comparisons with experimental activation energy data. All results presented use CHGNet–v0.3.0, unless noted otherwise; the bulk of the molecular dynamics production runs reported here were completed prior to the release of CHGNet–$r^2$SCAN and the development of the CHGNet–tuned model. Selected validation simulations with CHGNet–tuned are reported below to confirm that the qualitative trends identified with CHGNet–v0.3.0 are preserved under the higher-fidelity model.

## 3.2. Diffusion trends across composition and vacancy content

Molecular dynamics simulations in Figure 4 reveal a complex interplay between cation composition, oxygen vacancy concentration, and ionic transport in RE-HEOs; each data point represents the average of two independent simulations differing only in the initial placement of oxygen vacancies. At fixed Ce content, oxygen conductivity decreases with increasing vacancy concentration for the 22%, 31%, and 50% Ce compositions, as illustrated by the Arrhenius plots in Figure 4, with ionic conductivity as a function of temperature for a range of compositions and vacancy levels. This trend reflects a well-established phenomenon in aliovalent doped ceria: where excess oxygen vacancies begin to interact, forming clusters or trapping configurations that impede long-range migration [51,52]. As the vacancy concentration increases, percolation bottlenecks and reduced mobility emerge due to localized structural disruption. Interestingly, for the RE-HEOs this trend is not universal across compositions. In the 81% Ce system, the 3.13% oxygen vacancy exhibits both a lower activation energy and lower conductivity than the system with 9.38% vacancies. This observation deviates from the expected correlation between lower activation energy and higher diffusivity. It suggests that while the energy barrier for individual hops may be reduced, long-range transport is still hindered at lower vacancy content due to the limited number of available hoping sites. In this case, the increase in vacancy concentration at 9.38% appears to enable more extended diffusion pathways, despite potentially introducing some local trapping.

An analogous decoupling between activation energy and conductivity is well established in simpler RE-doped ceria fluorites (e.g., Gd-, Sm-, Y-, and Pr-doped $CeO_2$), where ionic conductivity exhibits a non-monotonic dependence on dopant, and therefore vacancy, concentration with a maximum near 10–20 mol% [6,9,15]. Below this maximum, transport is limited by the small number of mobile carriers despite low effective migration barriers, while above it, vacancy–vacancy repulsion and dopant–vacancy trapping dominate [6,15]. The 81% Ce RE-HEO behavior reported here sits on the low-vacancy (carrier-limited) side of this curve, indicating that the same balance between carrier availability and trapping governs transport in chemically complex RE-HEOs as in their binary doped-ceria analogs. These results emphasize the optimal vacancy content for diffusion is sensitive to the cation configuration and cannot necessarily be generalized across compositions.

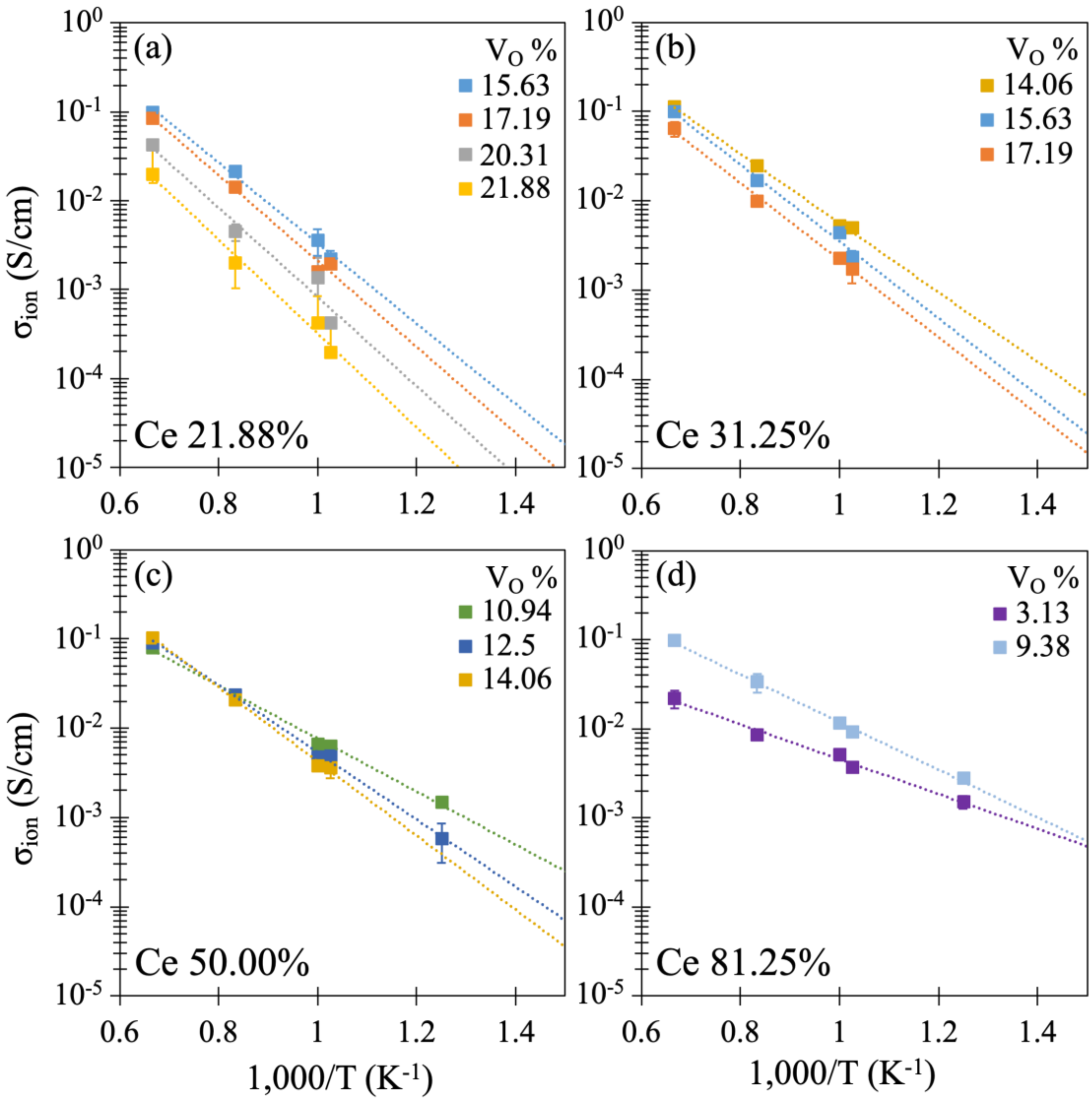

Figure 4. Conductivity vs. temperature for different $Ce_x(YLaPrSm)_{1-x}O_{2-\delta}$ systems from fluorite-initialized simulations. The dotted line is the linear fit that determines the activation energy. The error bars are the standard deviation between simulations at a given composition.

To evaluate whether finite-size effects influence the observed diffusion trends, additional MD simulations were performed using larger 4×4×4 supercells. Figure 5(b) and Figure 5(d) present the ionic conductivity as a function of temperature for the Ce 20% and Ce 50% compositions modeled with CHGNet–v0.3.0. The results reproduce the same general trends illustrated in Figure 4, namely that, at fixed Ce content, the oxygen ionic conductivity decreases with increasing oxygen vacancy concentration for both the 22% and 50% Ce compositions. Notably, for the equimolar 20% Ce composition, the inclusion of additional vacancy concentrations in the larger 4×4×4 supercell confirms that at low vacancy levels (2–10% $V_O$), ionic conductivity increases with increasing $V_O$, whereas beyond approximately 10% $V_O$, the conductivity decreases as the vacancy concentration rises to 20% $V_O$.

In Figure 4(c) and Figure 4(f), we also performed the same simulations for the smaller 2×2×2 supercells that were used in Figure 4(a) and Figure 4(d) but instead employing the fine-tuned CHGNet model. These results exhibit the same qualitative trend, ionic conductivity decreases at vacancy concentrations above 10%, yet the absolute conductivity values predicted by CHGNet–tuned are substantially higher than those from CHGNet–v0.3.0. This enhancement likely reflects the improved description of local structural relaxation and vacancy migration energetics achieved through fine-tuning to higher-fidelity r$^2$SCAN DFT data.

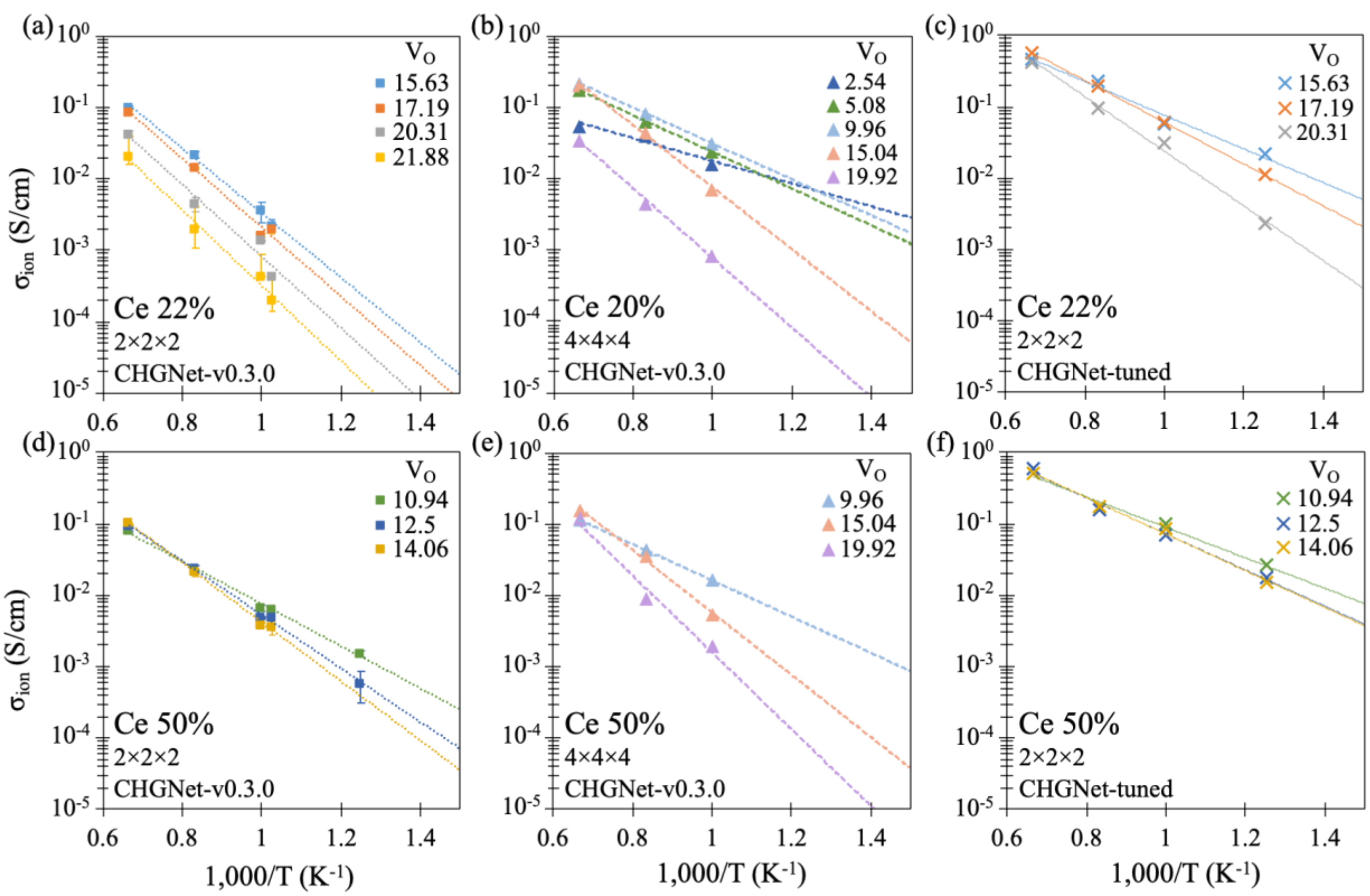

Figure 5. Conductivity vs. temperature for equimolar (a)-(c) and (d)-(f) 50% Ce content RE-HEO systems from fluorite-initialized simulations using either ((a), (b), (d), (e)) CHGNet-v0.3.0 or ((c),(f)) CHGNet-tuned MLIP with either ((a), (c), (d), (f)) 2x2x2 or ((b),(c)) supercells. Each line represents a oxygen vacancy composition and the dotted line is the linear fit that determines the activation energy. The error bars are the standard deviation between simulations at a given composition. For (b), (c), (e), and (f), only one simulation was run for a given data point.

Cation composition itself plays a significant role in dictating transport behavior. As shown in Figure 6(a) and Figure 6(b), increasing Ce content generally results in lower activation energies $E_a^{ion}$ and improved diffusivity. The activation energies in Figure 6(a) are obtained by first fitting an Arrhenius slope to the simulated conductivity–temperature data at each (Ce content, oxygen vacancy concentration) pair to extract a per-vacancy $E_a^{ion}$, and then averaging these per-vacancy $E_a^{ion}$ values across the range of oxygen vacancy concentrations simulated at fixed Ce content; the error bars are the corresponding standard deviations. This presentation isolates the Ce-dependence of the migration barrier from the independently characterized vacancy-concentration dependence shown in Figure 6(b),(d). This trend may be attributed to a combination of factors, including the lower average cation size variance, reduced lattice strain, and an increased likelihood of forming percolating Ce–Ce pathways that facilitate diffusion. For instance, in the Ce 81% composition, the diffusivity at 9% oxygen vacancy concentration is substantially higher than at 3%, reinforcing the importance of matching vacancy content to the local cation environment. In contrast, for Ce 22%, increasing the vacancy content from 15% to 22% leads to a reduction in conductivity, suggesting a shift beyond the optimal vacancy threshold into a regime dominated by trapping and percolation limitations.

The simulated $E_a^{ion}$ trend with Ce content shown in Figure 6(a) is the reverse of the experimental $E_a^{tot}$ values reported for the same and similar $Ce_x(YLaPrSm)_{1-x}O_{2-\delta}$ compositions [28], where $E_a^{tot}$ increases with increasing Ce content. This apparent disagreement reflects, at least in part, the different physical quantities being compared: the simulated $E_a^{ion}$ values report the activation energy for oxygen-ion migration in isolation, whereas the experimental $E_a^{tot}$ values are extracted from total conductivity measurements on samples that are known to be mixed ionic–electronic conductors, with Pr-mediated electronic conduction contributing increasingly as Ce content decreases. The experimental $E_a^{tot}$ therefore convolutes the ionic migration barrier with the temperature dependence of the electronic carrier concentration and mobility, while the simulations isolate the former. Given this difference in what each quantity represents, quantitative agreement between simulation and experiment for the magnitude or sign of the Ce-dependence trend should not be expected; what can be meaningfully compared is the order of magnitude, which is consistent across both. Structural phase effects were also explored in Figure 6(a). While both fluorite- and bixbyite-initialized configurations support oxygen migration, fluorite structures tend to exhibit slightly higher mobility at moderate to high temperatures, especially in Ce-rich systems. This may be due to the fluorite's more symmetric coordination environments, which support smoother migration pathways. However, in equimolar compositions, bixbyite structures display lower activation energies than fluorite, suggesting that local symmetry breaking and cation disorder may be in some cases promoting diffusion by increasing the diversity of accessible migration pathways.

Similar activation energy $E_{\mathrm{a}}^{\mathrm{ion}}$ trends were observed for both the larger 4×4×4 supercells using CHGNet-v0.3.0 and the smaller 2×2×2 supercells using CHGNet-tuned, as shown in Figure 6(c) and Figure 6(d). In both cases, the Ce 50% composition exhibits a lower $E_{\mathrm{a}}^{\mathrm{ion}}$ compared to Ce 20%, consistent with the enhanced ionic transport observed at higher Ce concentrations. As noted above for Figure 6(a), this simulated trend is opposite in sign to the experimental $E_{\mathrm{a}}^{\mathrm{tot}}$ reported for the same compositions; we attribute the reversal to the distinct quantities being measured ($E_{\mathrm{a}}^{\mathrm{ion}}$ in simulation versus $E_{\mathrm{a}}^{\mathrm{tot}}$ in experiment) rather than to a finite-size or potential-fidelity artifact, since both the larger 4×4×4 CHGNet–v0.3.0 and smaller 2×2×2 CHGNet–tuned results reproduce similar Ce-dependence. Additionally, Figure 6(d) shows that increasing oxygen vacancy concentration $V_O$ systematically increases $E_{\mathrm{a}}^{\mathrm{ion}}$, regardless of Ce content, indicating that excessive vacancy formation hinders long-range oxygen migration. The $E_{\mathrm{a}}^{\mathrm{ion}}$ values obtained from the larger supercells are comparable to those from the smaller ones, suggesting that finite-size effects are minimal. Notably, however, the absolute $E_{\mathrm{a}}^{\mathrm{ion}}$ values predicted by the CHGNet-tuned model are substantially lower than those obtained with CHGNet-v0.3.0, indicating the stronger diffusion behavior predicted by the tuned potential.

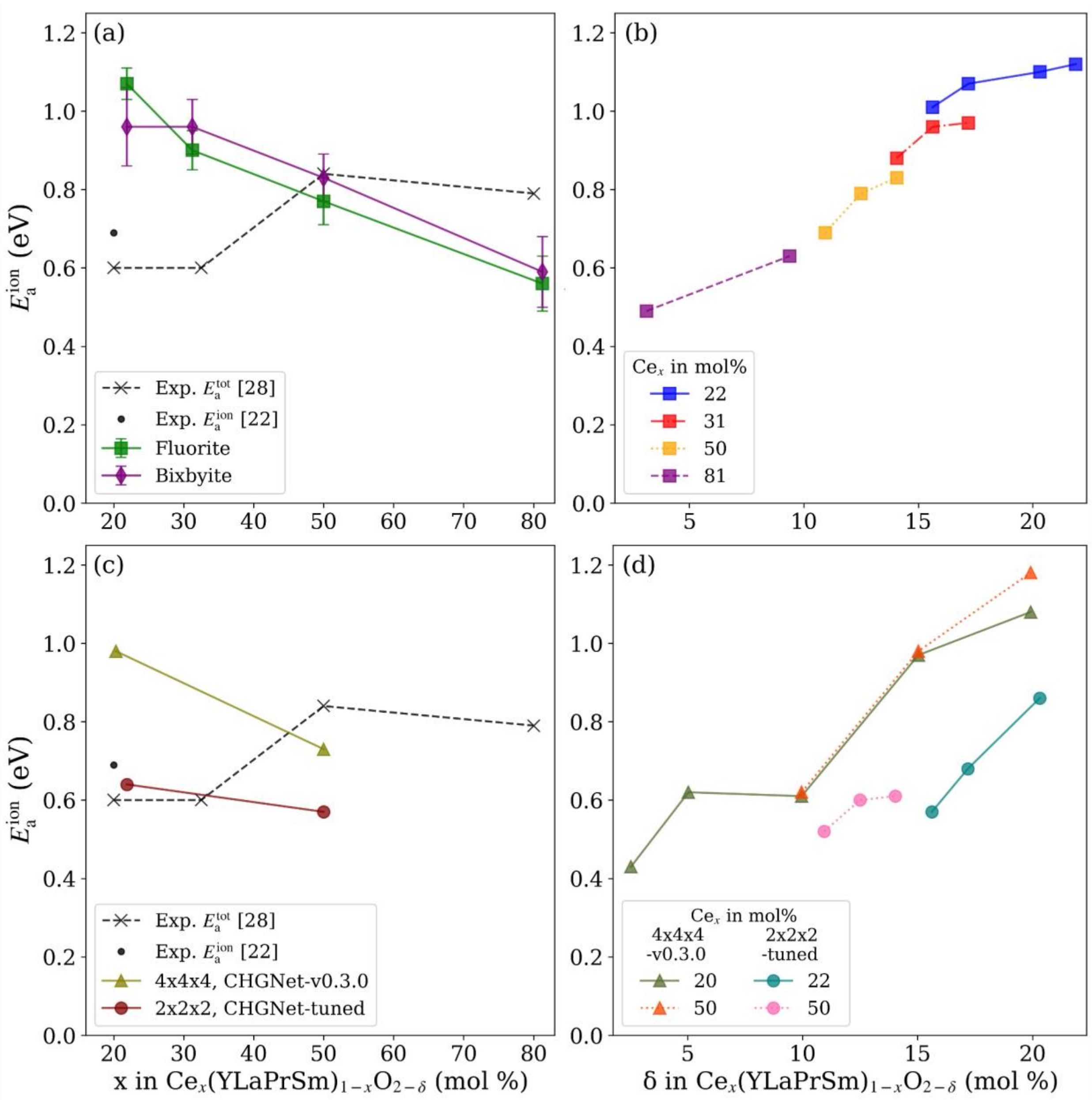

Figure 6. Activation energies $E_a$ for oxygen diffusion obtained from MD simulations as a function of (a),(c) Ce concentration and (b),(d) oxygen vacancy content. (a) $E_a$ as a function of Ce concentration, averaged over all oxygen vacancy concentrations listed in Figure 4, for fluorite-initialized (green) and bixbyite-initialized (purple) structures. (b)-(d) $E_a$ for the fluorite-initialized structures only. (c)-(d) $E_a$ from MD simulations using 4x4x4 supercells with CHGNet-v0.3.0 (triangles) or 2x2x2 supercells with CHGNet-tuned (circles). For the 4x4x4 supercells in (c) the $E_a$ for Ce 20% and Ce 50% were averaged over 15%-20% and 10%-15% $V_O$ concentrations, respectively. Experimental values taken from [28] and [22].

These results highlight the sensitivity of predicted transport properties to the underlying training data and model parametrization, while reaffirming the potential of machine-learned interatomic potentials for studying complex oxides. The CHGNet–tuned model offers improved agreement with experimental activation energies, suggesting a promising direction for future refinement and large-scale simulations aimed at bridging the gap between atomistic modeling and experimental measurements.

## 3.3. Microscopic mechanisms of oxygen transport

To further understand the atomistic origins of the predicted diffusion trends, we first visualized trajectory pathways using the OVITO software [53]. Figure 7 illustrates a representative diffusion trajectory from a 5ns MD simulation at 1200K for a composition consisting of 22% Ce content with 15% $V_O$. The red paths represent mobile oxygen ions, which exhibit hopping between tetrahedrally coordinated sites, while the blue paths correspond to immobile cations. The oxygen motion appears spatially homogenous, consistent with collective hopping dynamics across the lattice. Inspection of the oxygen trajectories reveals predominantly tetrahedral-to-tetrahedral hops between vacant sites, consistent with the expected vacancy-mediated diffusion mechanism. A small subset of trajectories appears as diagonal lines spanning non-adjacent sites; closer inspection of the trajectories shows that these correspond to correlated exchange events in which an oxygen hop into a vacancy is accompanied by a neighboring oxygen filling the newly vacated site within the same sampling interval. The diagonal appearance is therefore an artifact of the finite trajectory-output frequency rather than a physical long-range hop, and is consistent with concerted migration mechanisms previously reported in oxide ion conductors [54–56].

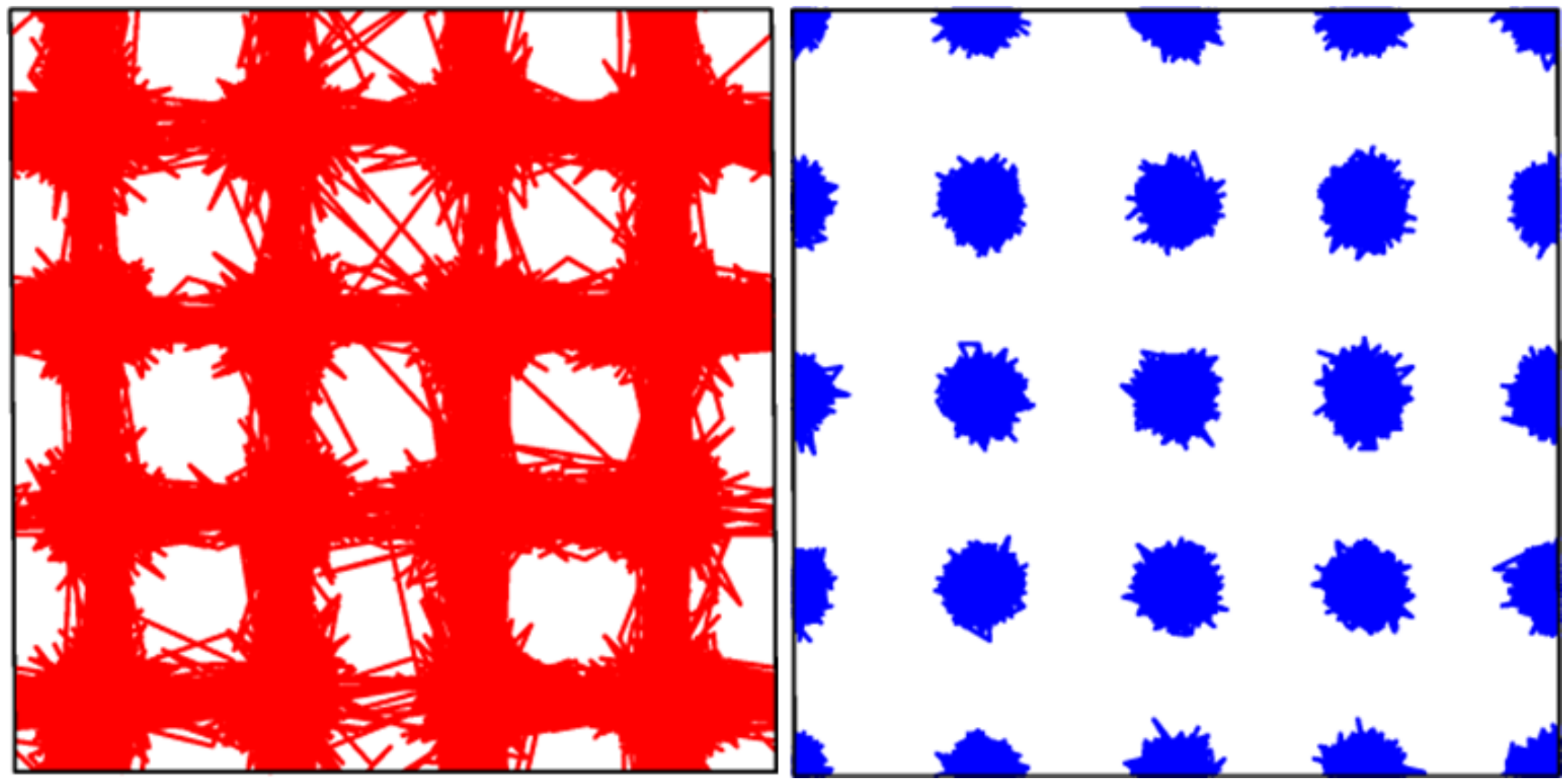

Figure 7. (left) Anion and (right) cation 5 ns trajectory lines for an equimolar fluorite with 15% Vo at 1200K. A small number of anion trajectories appear as diagonal lines connecting non-neighboring oxygen sites. These are visualization artifacts arising from finite frame sampling of concerted exchange events, in which one oxygen hops into a vacancy while a neighboring oxygen simultaneously occupies the just-vacated site, rather than genuine long-range diagonal hops.

To quantify these diffusion events, individual oxygen hops were identified and mapped to the specific cation–cation (RE–RE) edge through which each hop traversed. Figure 8 compares the normalized cumulative number of oxygen hops per RE–RE edge type for the two compositional endpoints of the $Ce_x(YLaPrSm)_{1-x}O_{2-\delta}$ series, Ce 22% (near-equimolar, maximum configurational entropy) and Ce 81% (Ce-dominated, minimum configurational entropy), at 1200 K. This pairing isolates the effect of configurational entropy on the hopping environment by holding the underlying chemistry fixed while spanning the accessible range of RE disorder; analogous distributions for the intermediate Ce 31% and Ce 50% compositions are provided in the Supplementary Information and follow the same qualitative trend. The proportion of oxygen hops occurring through Ce–Ce edges increase with Ce content. These Ce-rich edge environments support lower-energy transport channels, contributing to the observed reduction in activation energy and higher diffusivity. This compositional dependence on local transport environment highlights the importance of both global cation composition and short-range ordering in determining ionic transport behavior. With increasing Ce concentration, a greater fraction of hops occur through Ce–Ce and Ce–Y edges, contributing to reduced $E_a$ and higher diffusivity. These environments consistently support low-energy oxygen migration, likely due to the similar ionic radii of $Ce^{4+}$ (0.97 Å) and $Y^{3+}$ (1.019 Å), their relatively weak association to oxygen, and the formation of locally favorable lattice geometries. This interpretation is further supported by prior calculations showing comparable migration energies across Ce–Ce and Ce–Y edges [16,17,45,57–64]. These results provide microscopic evidence for Ce-dominated diffusion pathways in RE-HEOs.

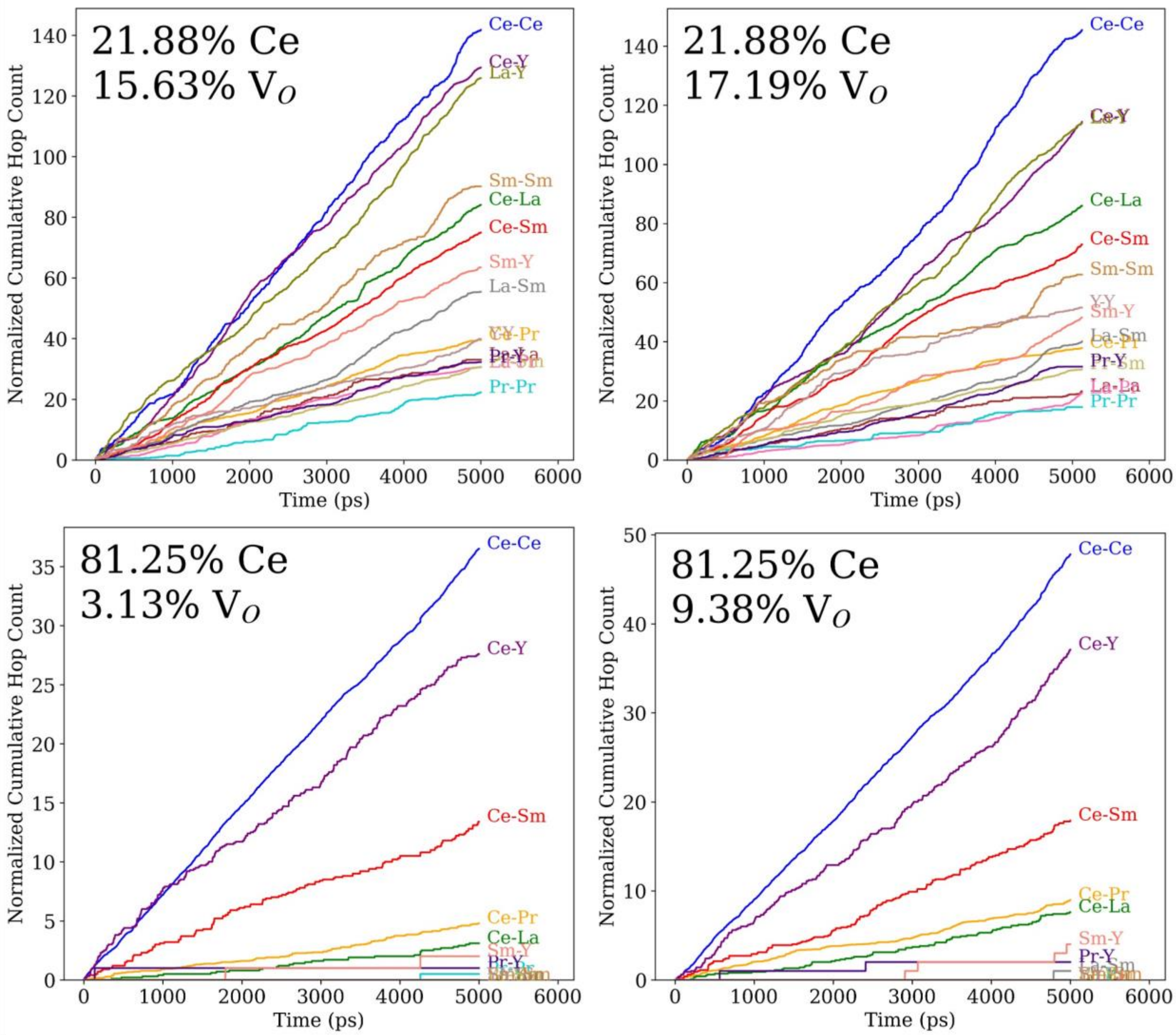

Figure 8. Normalized counts of oxygen ion hopping events through rare-earth (RE)–RE edge pairs at 1200 K for Ce 21.88% (top row) and Ce 81.25% (bottom row), with counts divided by the number of corresponding edge types present in the SQS structure. This normalization accounts for the varying abundance of different RE–RE edges, enabling a direct comparison of their relative contributions to oxygen diffusion pathways. Distributions for Ce 31% and Ce 50% are provided in the Supplementary Information.

The prominent role of Y in facilitating oxygen transport is further supported by coordination number analysis, which shows that Y has the lowest average oxygen coordination among all RE cations [Figure S.4]. This suggests that oxygen vacancies tend to localize near Y, increasing the likelihood of hops across Y-containing edges. Literature also reports favorable Y–$V_O$ association energies [16,57–60,62–64], reinforcing the mechanistic link between Y presence and enhanced oxygen mobility. Taken together, the data suggests that both Ce and Y play central, synergistic roles in shaping the local transport network within RE-HEOs. In contrast, edges involving Pr, Sm, and La are underrepresented in frequent hopping pathways. These cations are larger, introducing local strain and structural distortion that hinder hopping. In equimolar compositions, the broad distribution of cation sizes increases local lattice disorder, creating more frequent trapping sites and impeding long-range diffusion. Conversely, Ce-rich systems benefit from more uniform cation environments, reducing lattice strain and promoting smoother migration.

To systematically assess the transport activity of different cation–cation pathways, each RE–RE edge was classified based on how frequently it appeared among the five most and least active hopping pathways at 1200K across different Ce concentrations from both fluorite and bixbyite initialized systems. For each Ce concentration (22%, 31%, and 50%), the relative frequency of each edge's inclusion in the top or bottom five by normalized cumulative hop count is shown in Figure 9. Ce–RE and Y–RE edges are consistently among the most active diffusion pathways, whereas Pr–RE, Sm–RE, and La–RE edges appear disproportionately in the bottom five. These trends suggest that CHGNet captures chemically relevant trends in local transport behavior, identifying Ce and Y as key species for enhancing oxygen mobility in RE-HEOs.

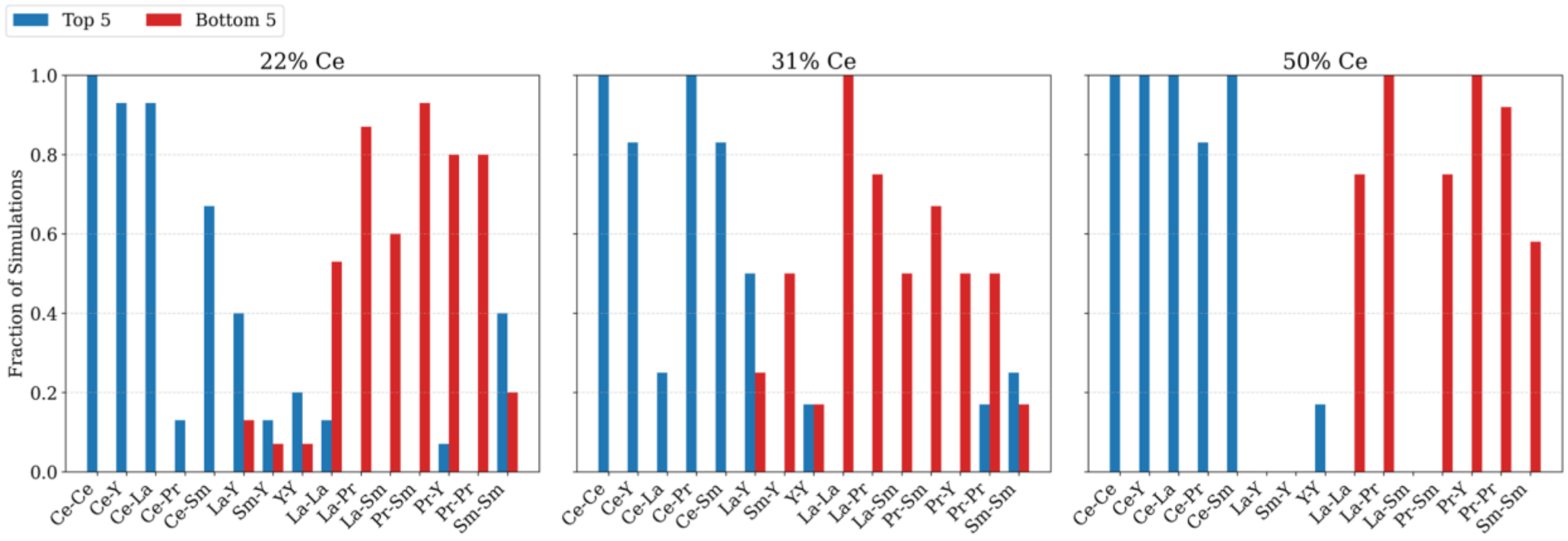

Figure 9. Frequency of RE–RE edges appearing in the top 5 and bottom 5 oxygen hopping pathways, as determined by cumulative hop counts across all simulations at 1200K. Results are shown for three different Ce concentrations: (left) 22%, (middle) 31%, and (right) 50%. Each bar represents the fraction of simulations in which the corresponding edge was ranked among the top 5 (blue) or bottom 5 (red) most active diffusion pathways. Ce–RE and Y–RE edges dominate the high-activity group across all compositions, while La-RE, Pr–RE, and Sm–RE edges are typically inactive.

# 4. CONCLUSIONS

The work described here demonstrates the effectiveness of a machine-learned interatomic potential, CHGNet, for modeling oxygen ion diffusion in RE-HEOs. Trends in ionic transport were revealed using molecular dynamics simulations on SQS-derived structures across a range of compositions and vacancy concentrations. Our results show that oxygen mobility in RE-HEOs depends critically on the interplay between vacancy concentration, cation composition, and the local structural environment. At low to moderate oxygen vacancy levels (~10–20%), increased vacant sites enhances diffusion by enabling more frequent hopping events. However, beyond this threshold, vacancy–vacancy interactions and trapping effects reduce overall conductivity, echoing trends previously observed in doped $CeO_2$ systems. Increasing Ce content reduces the activation energy for diffusion and increases overall conductivity, particularly at higher temperatures. This enhancement is attributed to the emergence of more favorable diffusion pathways, dominated by Ce–Ce and Ce–Y cation edge connectivity, that support low-barrier oxygen migration. These results offer atomistic insight into how cation choice and distribution can be leveraged to engineer fast ion conductors. Despite CHGNet's lack of training on explicit *f*-electron treatment, the simulations reproduce good agreement with trends in experimental conductivity and provide a valuable atomistic perspective on ion transport. Although, the randomized, idealized distribution of oxygen vacancies may not capture possible defect clustering or short-range ordering, which can further influence transport in real materials, the model successfully captures dominant structure–transport relationships in fluorite RE-HEOs.

Overall, this study establishes machine-learned interatomic potentials as a promising tool for simulating complex oxides and accelerating the design of next-generation functional ceramics. RE-HEOs offer an exceptionally tunable platform for optimizing ionic conductivity via structural and chemical control of defects, and this work contributes foundational insights into their structure–diffusion relationships. Continued efforts to incorporate correlated defect behavior, multivalency, and longer timescale dynamics will further advance our understanding of these complex systems.

**CRediT authorship contribution statement:**

**Mary Kathleen Caucci**: Writing – original draft, Writing – review & editing, Visualization, Validation, Methodology, Investigation, Formal analysis, Data curation, Conceptualization, Software. **Billy Yang**: Writing – review & editing, Conceptualization. **Saeed Almishal**: Writing – review & editing, Conceptualization. **Jon-Paul Maria**: Writing – review & editing, Funding acquisition, Conceptualization. **Susan Sinnott**: Writing – review & editing, Supervision, Resources, Funding acquisition, Conceptualization.

**Data Availability:**

The DFT dataset used to fine-tune the CHGNet–tuned model and an analysis script used for oxygen-hop identification and rare-earth pair attribution are openly available at Open Science Framework under the DOI https://doi.org/10.17605/OSF.IO/W38CM. Molecular dynamics simulations were performed using the LAMMPS code, and the machine-learned interatomic potentials are based on the publicly available CHGNet models; both are openly available from their respective developers.

**Acknowledgements:**

The authors would like to thank Jacob T. Sivak for useful discussions. The authors acknowledge the use of facilities and instrumentation supported by NSF through the Pennsylvania State University Materials Research Science and Engineering Center [DMR-2011839]. Computations for this research were performed on the Pennsylvania State University's Institute for Computational and Data Sciences' Roar supercomputer.

# Supplementary Information

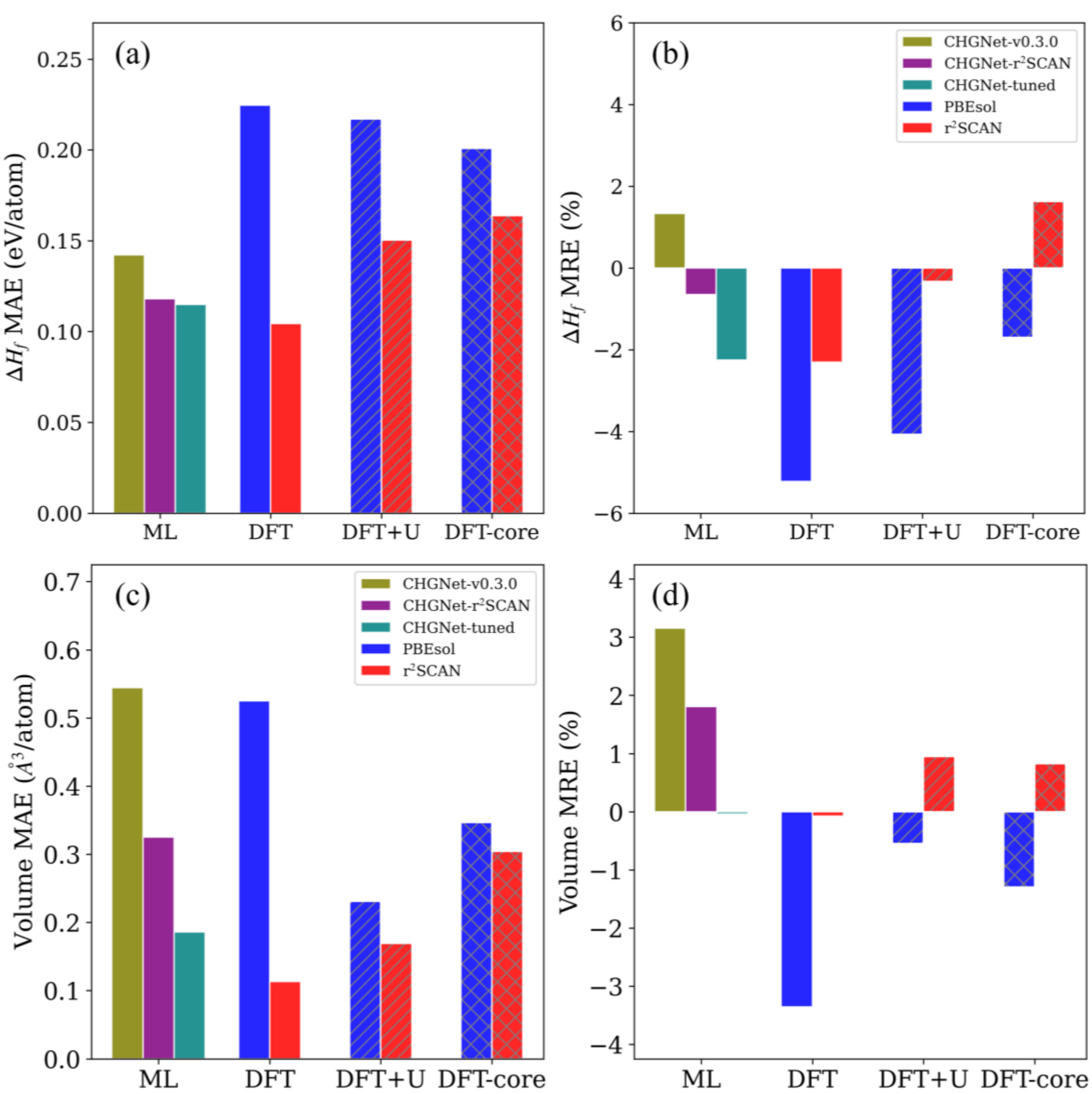

Figure S.1. The (a),(c) mean absolute errors, MAE, and the (b),(d) mean relative errors, MRE, of the MLIP, DFT, DFT+$U$, and DFT-core (a) – (b) formation enthalpies $\Delta H_f$ and (c) – (d) lattice volume predictions compared to experiment for 15 REOs using CHGNet (v0.3.0, $r^2$SCAN, tuned), PBEsol, and $r^2$SCAN. Note that DFT-core does not include predictions for $Y_2O_3$, $La_2O_3$, $PrO_2$, $CeO_2$. DFT results reproduced from Ref. [41].

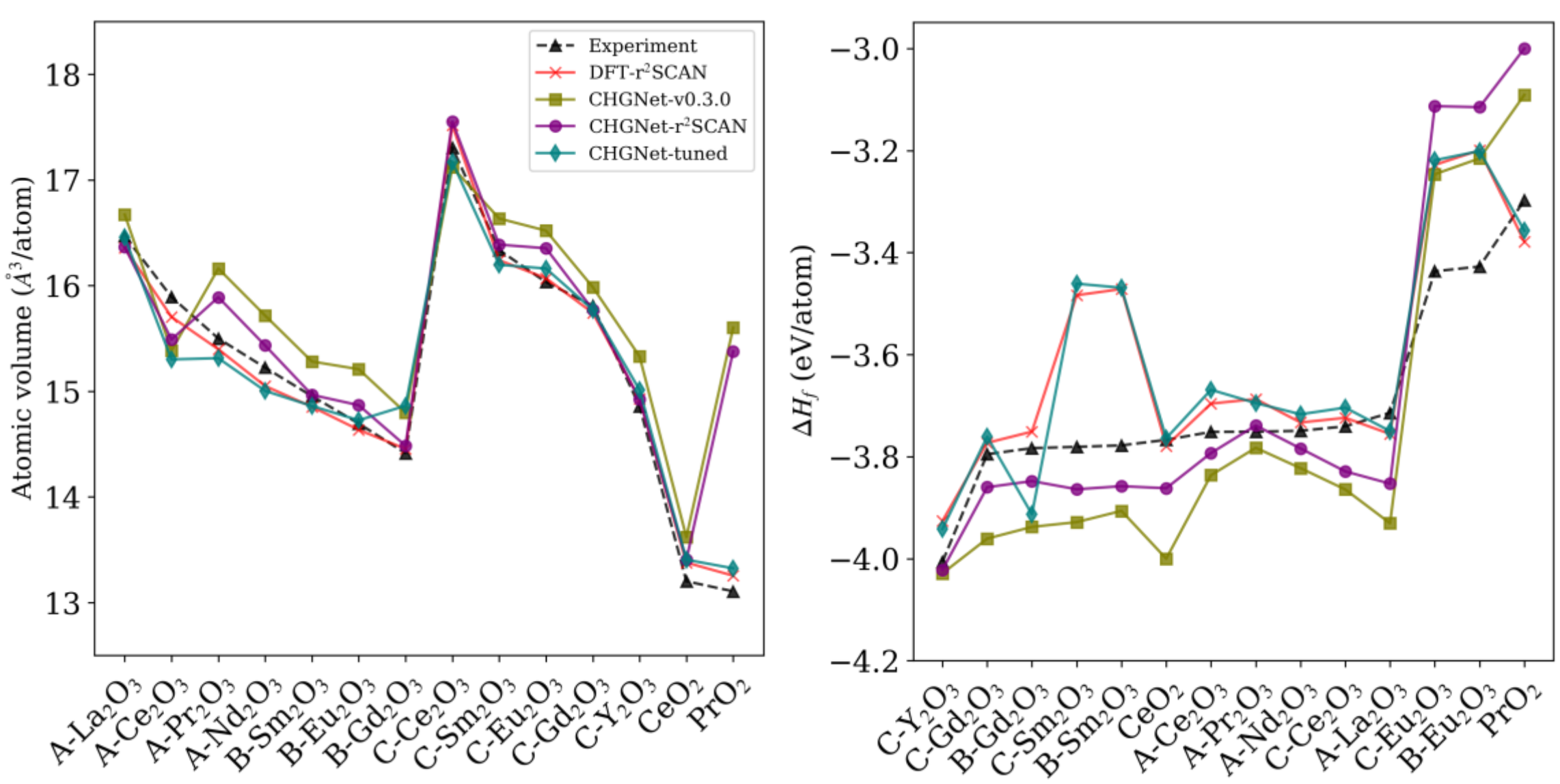

Figure S.2. The predicted (a) lattice volumes and (b) formation enthalpies for considered REOs using DFT--$r^2$SCAN (red cross), CHGNet-v0.3.0 (olive square), CHGNet-$r^2$SCAN (purple circle), and CHGNet-tuned (teal diamond) in comparison to experimental values (black triangle). DFT results reproduced from Ref. [41].

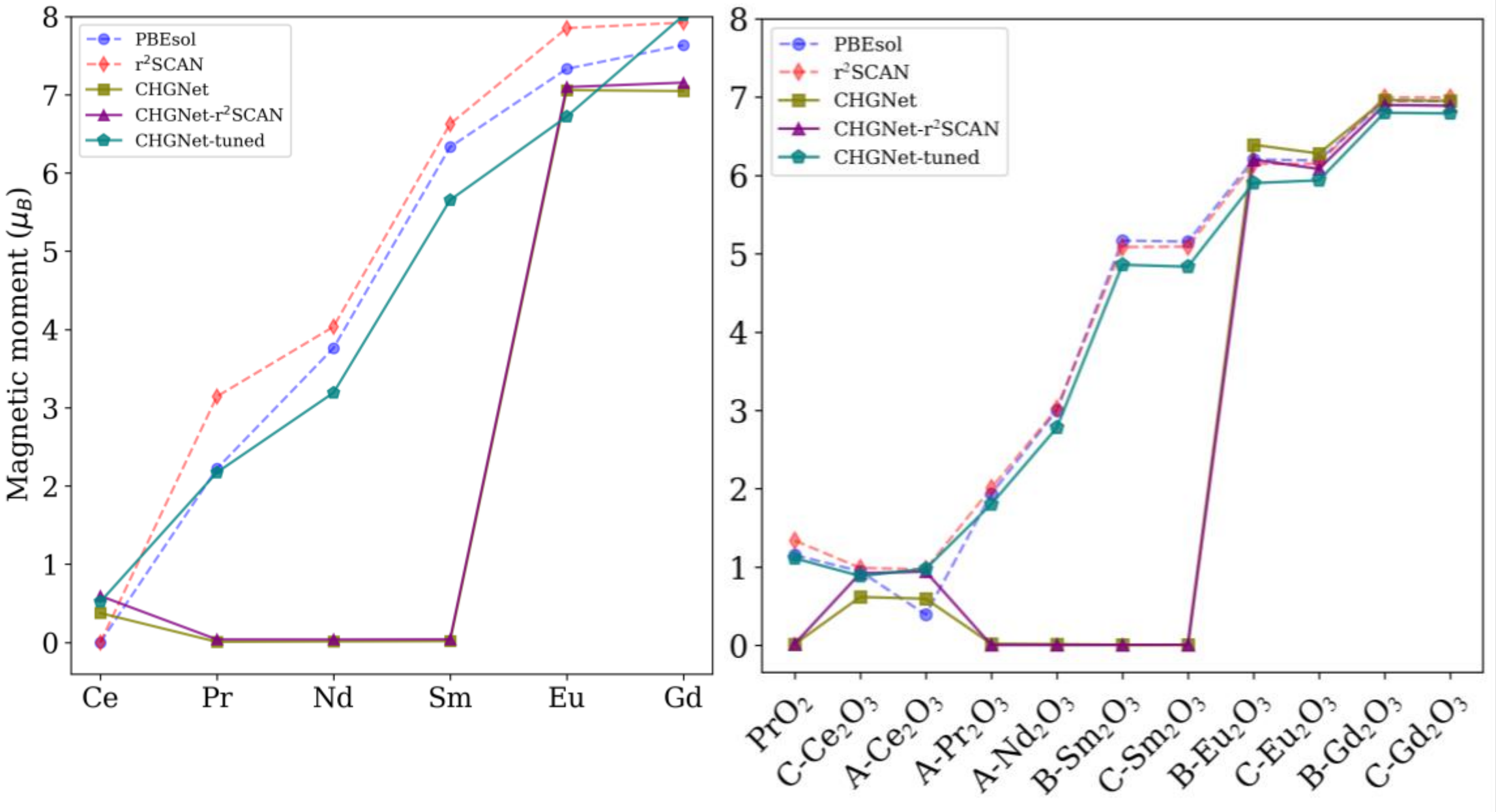

Figure S.3. The calculated average total magnetic moment (in μB) for the RE cations in either (a) metallic or (b) oxide phase using CHGNet-v0.3.0 (olive squares), CHGNet-$r^2$SCAN (purple triangles), CHGNet-tuned (teal pentagon), and DFT with either PBEsol (blue circles) or $r^2$SCAN (red diamonds) XC functionals. Solid lines are to guide the eye. DFT results reproduced from Ref. [41].

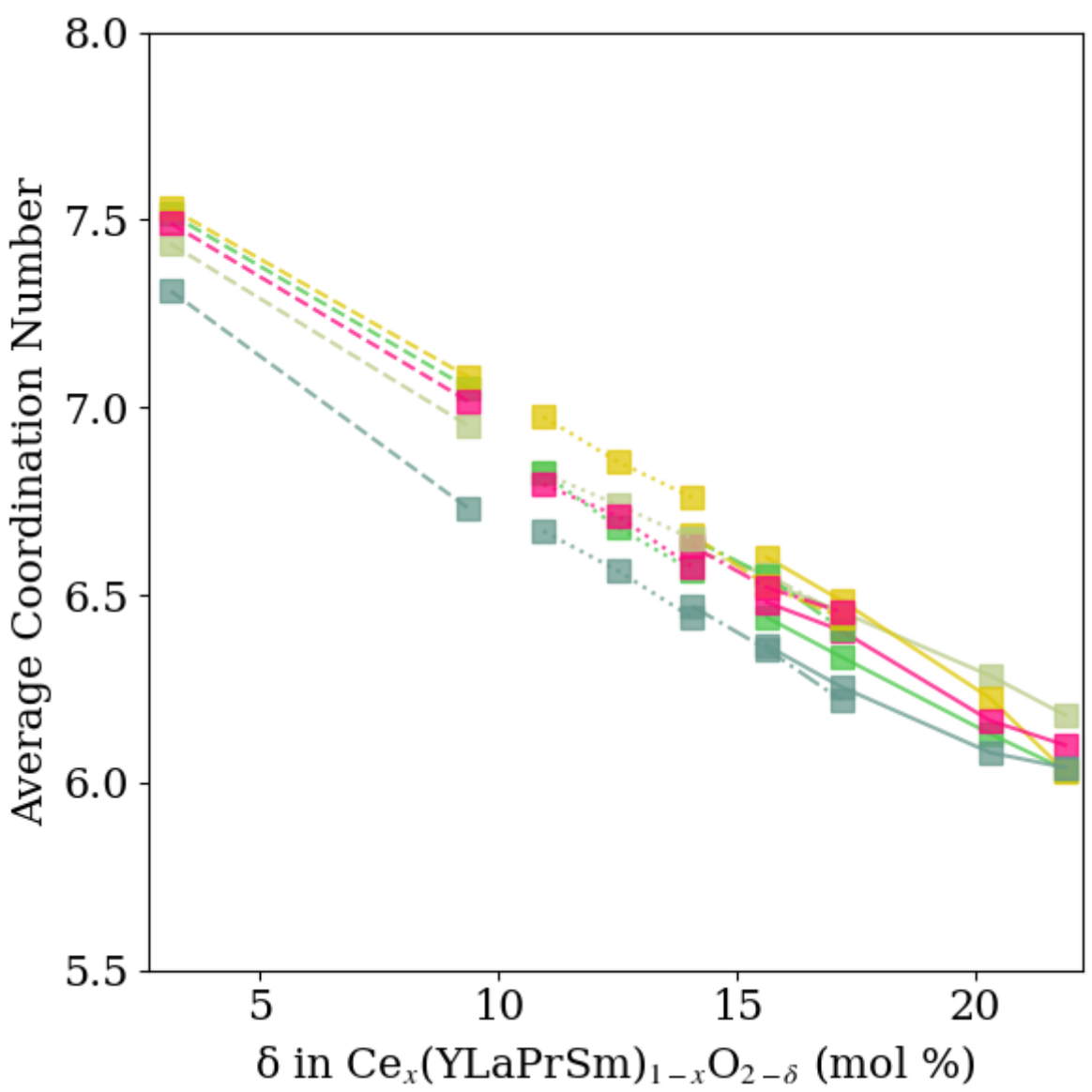

Figure S.4. Average coordination for RE elements from the 5ns simulations at 1200K for fluorite-initialized Ce 22% (solid line), Ce 31% (dot-dash line), Ce 50% (dotted line), and Ce 81% (dashed line).

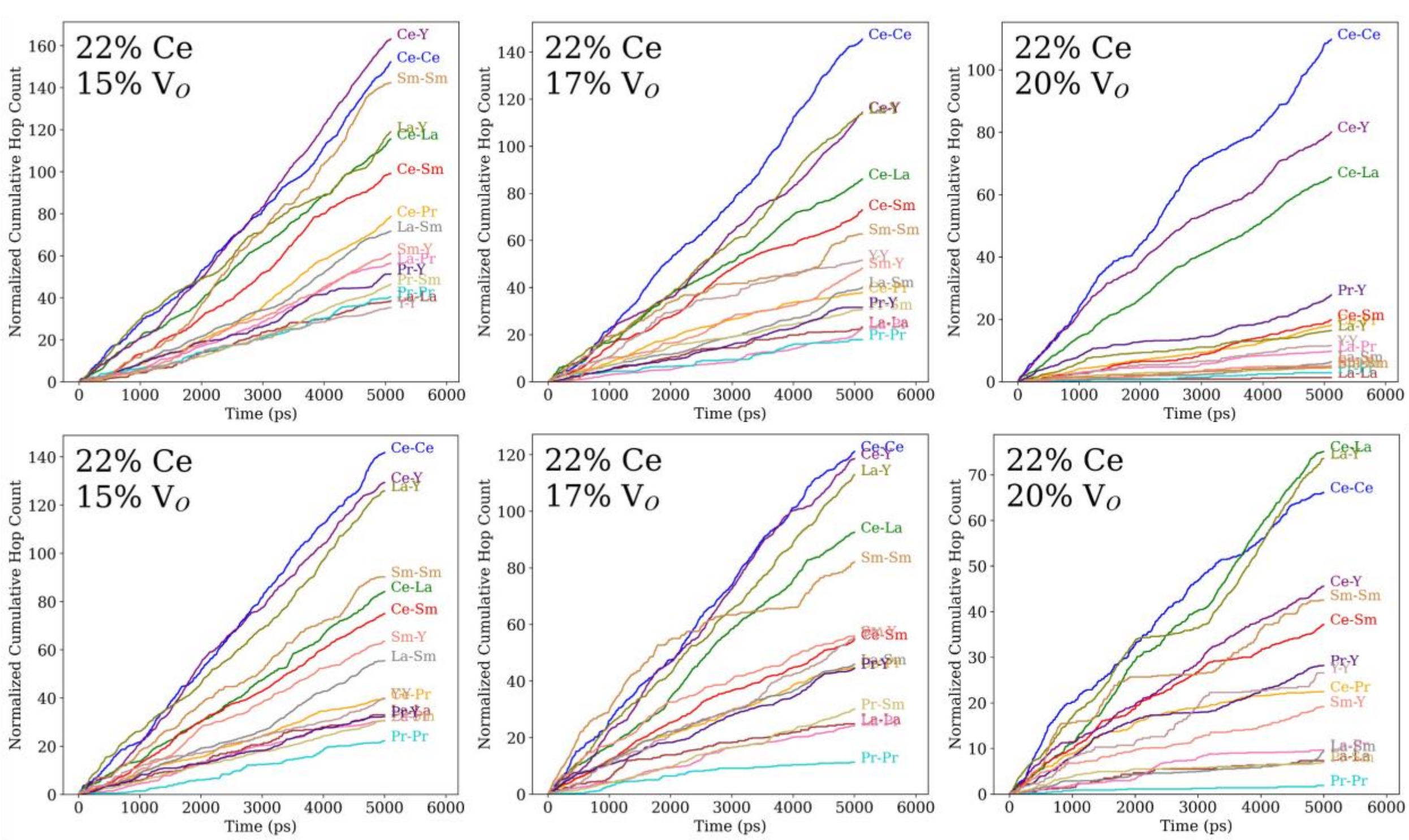

Figure S.5. Normalized counts of oxygen ion hopping events through rare-earth (RE)–RE edge pairs at 1200 K for each Ce 21.88% fluorite initialized simulation, with counts divided by the number of corresponding edge types present in the SQS structure. Each vacancy result differs only in the initial placement of oxygen vacancies.

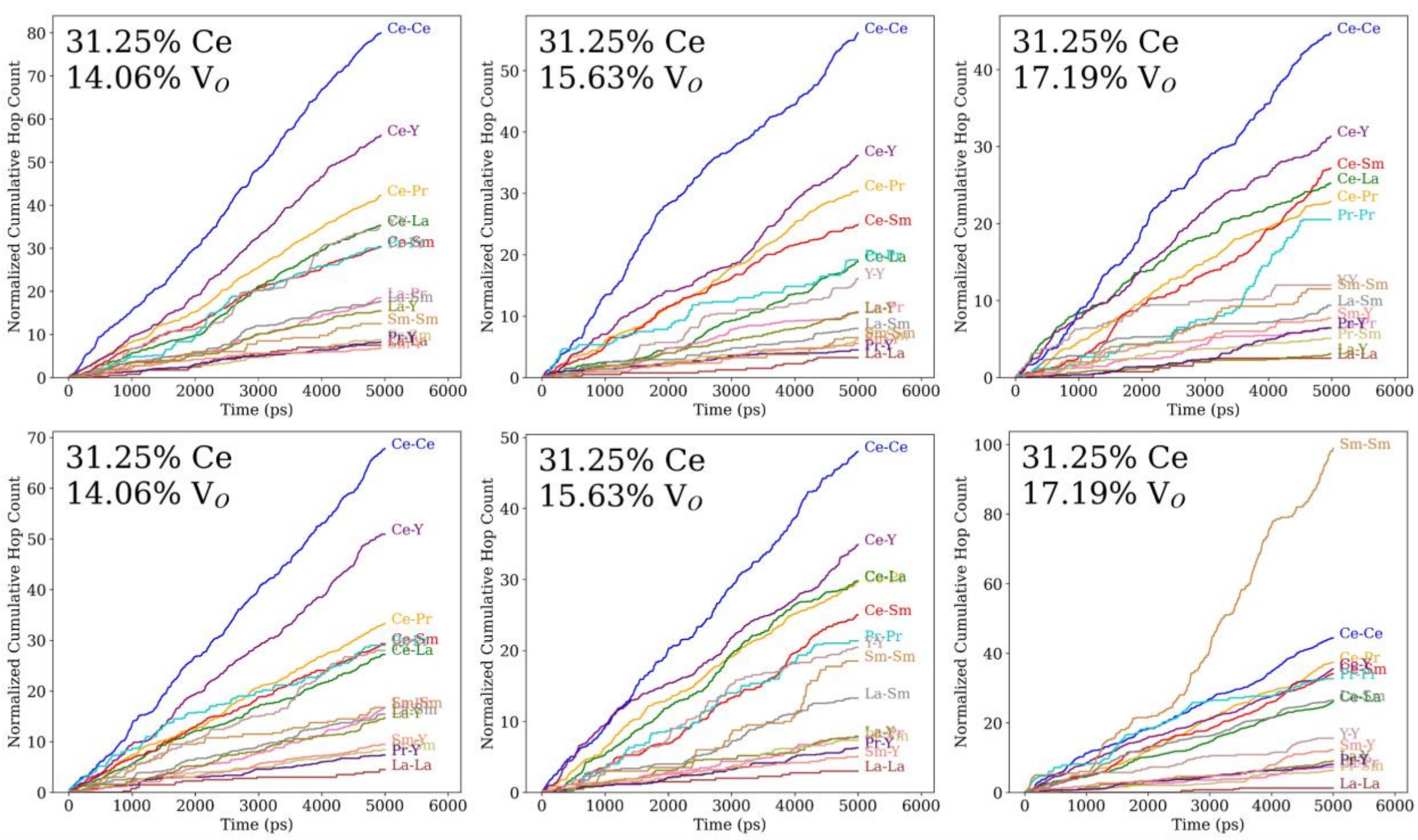

Figure S.6. Normalized counts of oxygen ion hopping events through rare-earth (RE)–RE edge pairs at 1200 K for each Ce 31.25% fluorite initialized simulation, with counts divided by the number of corresponding edge types present in the SQS structure. Each vacancy result differs only in the initial placement of oxygen vacancies.

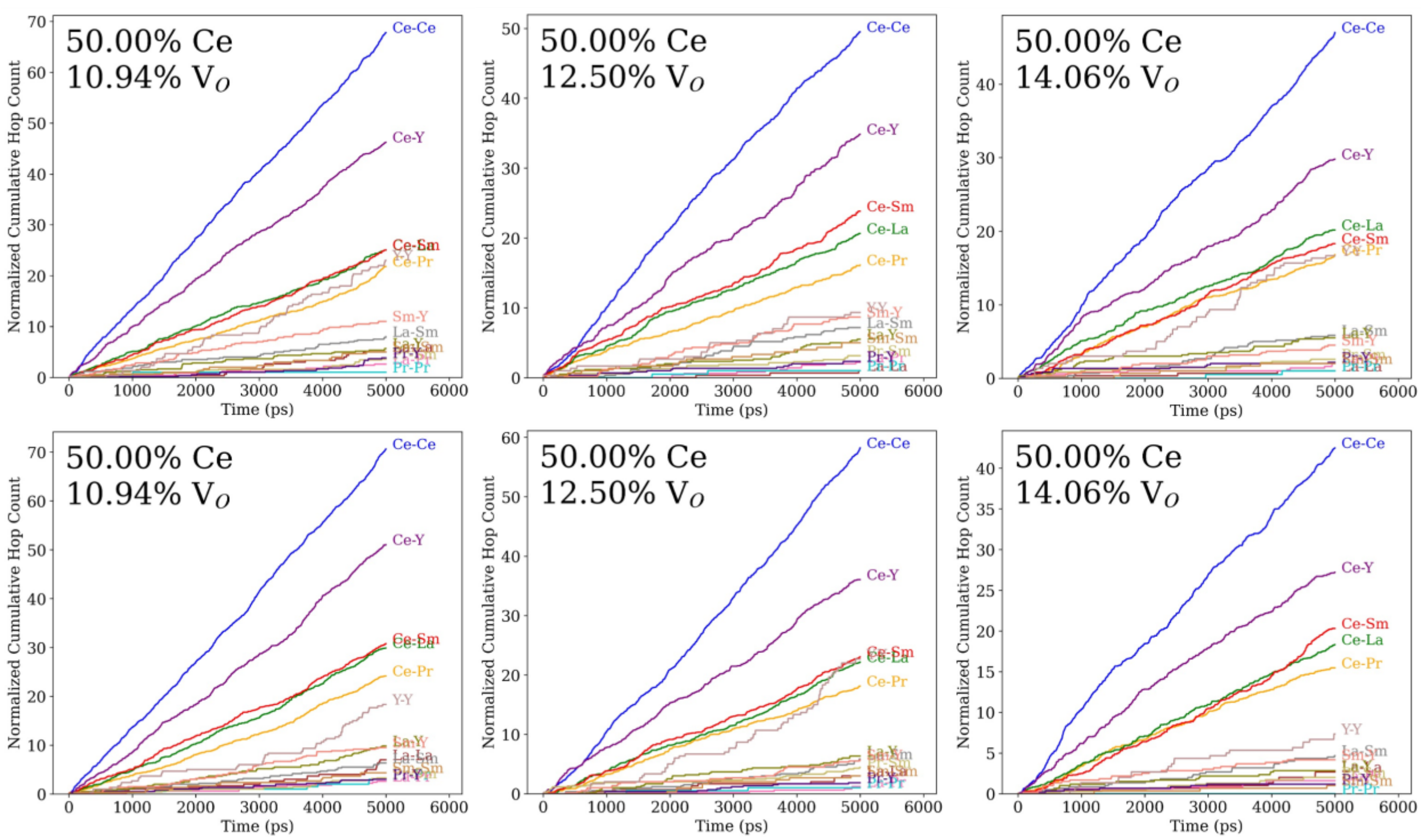

Figure S.7. Normalized counts of oxygen ion hopping events through rare-earth (RE)–RE edge pairs at 1200 K for each Ce 50% fluorite initialized simulation, with counts divided by the number of corresponding edge types present in the SQS structure. Each vacancy result differs only in the initial placement of oxygen vacancies.

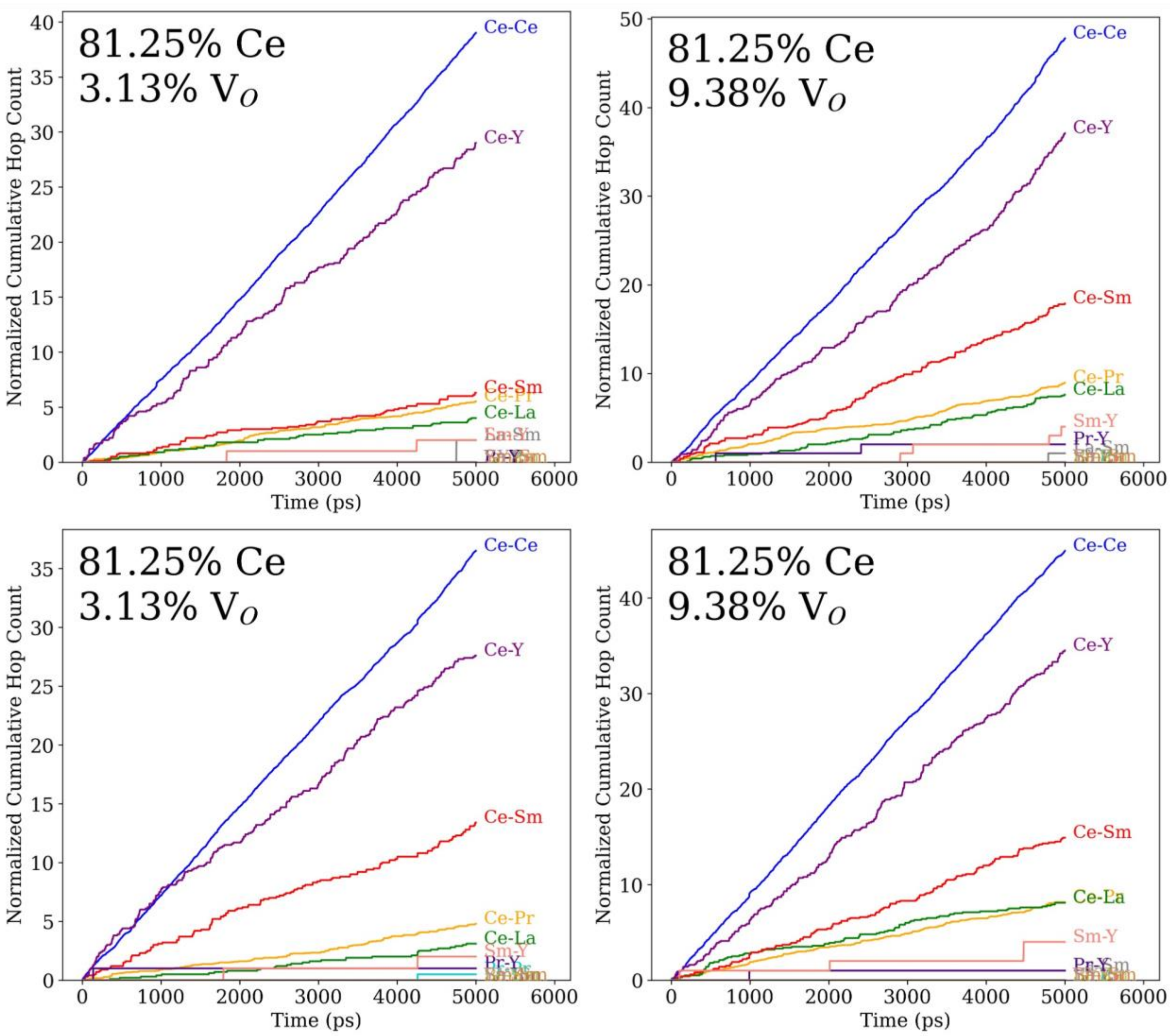

Figure S.8. Normalized counts of oxygen ion hopping events through rare-earth (RE)–RE edge pairs at 1200 K for each Ce 81.25% fluorite initialized simulation, with counts divided by the number of corresponding edge types present in the SQS structure. Each vacancy result differs only in the initial placement of oxygen vacancies.